\documentstyle[prb,aps,preprint]{revtex}
\tightenlines
\author{Simone Artz, Michael Schulz and Steffen Trimper}
\address{Fachbereich Physik\\Martin-Luther-Universit\"at\\D-06108 Halle\\
Germany}
\title{Diffusion with restrictions}
\begin{document}
\draft
\maketitle
\begin{abstract}
\noindent A non--linear diffusion equation is derived by taking into account 
hopping rates depending on the occupation of next neighbouring sites. 
There appears additonal repulsive and attractive forces leading to a changed 
local mobiltiy. The stationary and the time dependent 
behaviour of the system are studied based upon the master equation approach. 
Different to conventional diffusion it results a time 
dependent bump the position of which increases with time described by an 
anomalous diffusion exponent. The fractal dimension of this random walk is 
exclusively determined by the space dimension. The applicabilty of the model 
to descibe glasses is discussed.\\*[2cm]  

\end{abstract}

\noindent There has been a lot of effort  in understanding strongly interacting liquids, 
especially glass--forming, undercooled liquids \cite{go1,ja}. Obviously, 
the liquid--glass transition is primarily dynamic in origin characterized 
inevitably by a high cooperativity of local processes \cite{ag}. Due to 
the seminal papers of G\"otze et al \cite{go2} and independently of 
Leutheuser \cite{le} the mode coupling theory for dense liquids offers a 
theoretical tool to study the slow dynamics in the vicinity of the glass 
transition. The mode coupling theory implies that the (ergodic) 
liquid state can be characterized by the decay of correlations whereas 
below a critical temperature $T_c$ predicted by the approach it results a 
crossover to a solid--like behaviour where fluctuations are effectively 
frozen in. Furthermore, if additional thermal activated diffusion processes 
are taken into account the correlations decay  but on a longer time scale 
\cite{dm,go3,sj}. The extended mode coupling theory \cite{fr1,fr2} leads to 
a more realistic behaviour of the correlation function.\\
\noindent The present paper is motivated by another kind of approach to 
supercooled liquids and glasses. It is based on a proposal due to 
Fredrickson--Andersen (FA) \cite{fa} to map the system onto an Ising model 
where the spin variables are related to local regions with a lower or a higher 
density than an averaged one. Such a facilitated model is analyzed above all 
numerically \cite{fb} but recently also analytically \cite{1d,stmfa}. 
Different to the conventional kinetic Ising model the dynamical rules for 
spin-flip processes are changed in order to take into account the enhanced 
immobility of the particles due to the gradual freezing with decreasing 
temperature. These restrictions can be easily formulated in a Fock space 
representation of the master equation by assuming flip rates which are 
dependent on the occupation of neighbouring sites promoting or preventing the 
processes under consideration. In this manner the inherent cooperativity 
manifested in a cooperative rearrangement of certain regions in order to 
change mobile states within the subsequent time steps into immobile ones and 
vice versa. The inclusion of mobile vacancies in the facilitated model 
\cite{sr1,pi} yields to the occurence of nearly nonactivated relaxation 
additional to the main $\alpha$--process.\\
Here, we study another kind of secondary processes. They are due to the 
ability of the particles to diffuse when the system is above the so called 
glass temperature $T_g$. The diffusion constant goes to zero when the 
temperature is further decreased. Obviously, the mobility of the system 
is more and more reduced whenever the temperature reaches the glass 
temperature. Here, we consider diffusion for $T > T_g$ but take into account 
restrictions which lead to the above mentioned diminishing mobility. 
To this aim particles can jump from lattice site to lattice site which 
gives rise to diffusion on a large scale. To simulate the freezing 
process the hopping rate depends on the occupation of the nearest 
neighbour site of a given pair of sites between which a particle can jump. In such 
a manner the diffusion is more and more confined through the environment.\\ 
To describe this situation we use the mentioned Fock--space representation for 
the master equation under the influence of an exclusive dynamics. Such 
processes had been discussed in 
different contexts, compare for instance \cite{sp,gw,satr,scsa,aldr,li}. 
As already mentioned the starting point is the master equation which 
describes the stochastic process. A systematic approach is achieved by 
a mapping of the stochastic equation onto an evolution operator entering as 
the infinitesimal time evolution operator. This operator can be expressed in 
terms of second quantized operators. Physically it means the considered 
system is divided in small cells of size l (lattice points). 
Every cell is occupied by particles underlying dynamical rules depending on 
the situation in mind. From a quantum 
point of view two limiting cases can be realized, an unrestricted occupation 
number and a restricted one obeying the Pauli-principle with the occupation 
numbers 0 and 1. This situation corresponds to a lattice gas representation with 
empty and single occupied sites which can be also interpreted as the 
two orientations within a spin model.\\ 
The case of unrestricted occupation had been discussed firstly by Doi 
\cite{doi} and had been further elaborated in \cite{satr,gra,pe}. 
Here, we consider the case of a restricted occupation cell. Hence, the 
problem is to formulate the dynamics in such a way that this restriction is 
taken into account. The method offers a wide range of application, for a 
recent summary see \cite{sti}.\\ 

\noindent The starting point is the Master equation which can be derived on quite 
general grounds. It can be written in the form  
\begin{equation}
\partial_tP(\vec n,t)=L^{\prime}P(\vec n,t) 
\label{ma}
\end{equation}
where $P$ is the probability that a certain configuration characterized 
by the vector $\vec n$ at time $t$ is realized. The evolution operator 
$L^{\prime }$ will be specified by the dynamics of the model. Following 
\cite{satr,aldr,doi,gra,pe} the probability distribution 
$P(\vec n,t)$ can be related to a state vector $\mid F(t) \rangle$ in a 
Fock-space 
according to $P(\vec n,t) = \langle \vec n\mid F(t)\rangle$ with the 
basisvectors $\mid \vec n \rangle$. 
As a consequence the Master equation (\ref{ma}) can be transformed to an equivalent 
equation in a Fock-space
\begin{equation}
\partial_t \mid F(t)\rangle = \hat{L} \mid F(t) \rangle
\label{fo1}
\end{equation}
The operator $L'$ in (\ref{ma}) is mapped onto the operator $\hat{L}$ 
given in a second quantized form. Up to now the procedure is independent on 
the used operators. Usually $\hat{L}$ is expressed in terms of 
creation and annhiliation operators satisfying Bose commutation 
rules \cite{doi,gra,pe}. To avoid double occupancy as for instance 
in a forest fire model \cite{pt} or a model simulating traffic jam \cite{kt} 
the method can be extended to the case of restricted occupation numbers per 
lattice sites, \cite {gw,satr,scsa,aldr} introducing Pauli-operators. 
These operators commute for different points and anticommute at the same 
lattice point. A further extension to a p--fold occupation has been discussed 
recently \cite{st2,st3}\\
\noindent The relation between the quantum formalism and the probability approach 
based upon the Master equation can be found by expanding the vector 
$\mid F(t) \rangle$ with respect to the basisvectors of the Fock-space
\begin{equation}
\mid F(t) \rangle = \sum_{n_i} P(\vec n,t) \mid \vec n \rangle
\label{fo2}
\end{equation}  
As it was shown firstly by Doi \cite{doi} the average of an arbitrary physical 
quantity $B(\vec n)$ is given by the average of the corresponding operator 
$\hat{B}(t)$
\begin{equation}
\langle \hat{B}(t) \rangle = \sum_{n_i} P(\vec n,t) B(\vec n) = 
\langle s \mid \hat{B} \mid F(t) \rangle 
\label{fo3}
\end{equation} 
This rule remains also valid as well as in case of Pauli-operators 
\cite{satr} and parafermi-operators however with different meaning of the 
state vector $\mid s \rangle$ introduced in the last equation. 
Remark that the normalization condition for the state function is 
manifested in the relation 
$\langle s \mid F(t) \rangle = 1$ and the sum-vector $\langle s \mid$ can be 
expressed by $\langle s \mid = \sum \langle \vec n \mid$. 
The evolution equation for an operator $\hat{B}$ can be written  
\begin{equation}
\partial_t \langle \hat{B} \rangle = \langle s \mid [\hat{B},\hat{L}] \mid F(t) \rangle
\label{kin}
\end{equation}
To derive the last relation we have used $\langle s \mid \hat{L} = 0$. 
Note that dynamical equations of the classical problem are determined 
by the commutation rules of the underlying operators and the form of the 
evolution operator.\\

\noindent Now let us demonstrate the applicability of the method to describe 
diffusive motion under the influence of a constraint. In the second 
quantized formulation the hopping process is simply 
descibed by the evolution operator 
\begin{equation}
\hat{L} = \sum_{<ij>} J_{ij} 
\left(d^{\dagger}_i d_j - (1 - n_i) n_j \right)
\label{fli1}
\end{equation}
where the hopping rate $J_{ij}$ between the adjacent sites $i$ and $j$ 
can also depend on the occupation number of the next nearest sites, see below.  
Furthermore we have introduced the occupation number operator 
$$
n_i = d^{\dagger}_id_i
$$
Because the original problem is a classical one the operators 
$d_i$ and $d^{\dagger}_i$  satify the commutation rules for Pauli--operators. 
They commute at different lattice sites and anticommute at the same sites. 
$$
[d_i, d^{\dagger}_j] = \delta_{ij}(1 - 2 n_i)
$$
Using eq.(\ref{kin}) and the commutation relations we get the evolution 
equation for the averaged occupation number 
\begin{equation}
\partial_t \langle n_r \rangle = \sum_{r(j)} \langle J_{rj} (n_j - n_r) \rangle 
\label{fli2}
\end{equation}

\noindent The evolution operator (\ref{fli1}) allows diffusive processes 
between nearest neighbour sites whenever a double occupancy is avoided. 
Now let us generalize the model by including the above mentioned topological 
restrictions due to the freezing of the system. To this aim we have to 
include that the hopping rates $J_{rj}$ depend on the occupation number 
$n$.
As the simplest nontrivial assumption we chose
\begin{equation}
J_{rj} = J \sum_l \chi_{rjl} n_l
\label{restr}
\end{equation}
Here, $\chi_{rjl}$ is only nonzero when the triple of lattice indices $rjl$ 
denotes nearest neighbours. Introducing the abbreviation $\Theta_{ij} = 1$ 
when $i$ and $j$ are nearest neighbours and zero otherwise one finds
$$
\chi_{rjl} = \Theta_{rj} \Theta_{rl} (1 - \delta_{jl}) 
$$
Let us visualize the proceeses for the one dimensional case. A particle 
can jump from site $i+1$ to the empty site $i$ when the site $i-1$ is 
occupied. On the other hand a particle jumps from site $i$ to the empty 
site $i+1$ under the condition that the site $i-1$ is occupied. In such 
a manner the hopping process is performed under the influence of attractive 
and repulsive forces as described before.\\
The kinetic equation for the averaged density reads 
\begin{equation}
\partial_t \langle n_r \rangle = J \sum_{jl} \chi_{rjl} 
\langle n_j n_l - n_r n_l \rangle
\label{kin1}
\end{equation}
There occurs a hierarchy of evolution equations which cannot be solved 
in general.
Here  the higher order terms are decoupled due to a
mean field approximation. In the resulting equation  we perform the continuum limit 
for the field $n(\vec x, t) = \langle n_i(t) \rangle$. The corresponding evolution 
equation is expanded upto the order of $l^2$, where $l$ is the cell size introduced 
before. The corresponding evolution equation in this  
order  and in a low density expansion is written in the form 
\begin{equation}
J^{-1}\partial_t n = (z -1) n \nabla^2n - 2 (\nabla n)^2
\label{stat}
\end{equation}
where $z$ is the coordination number and for simplicity we have set $l=1$. 
In case of a simple cubic lattice 
$z$ is simply related to the space dimension via $z = 2d$. Because the hopping 
rate $J_{ij}$ depends on the concentration itself the dynamical equation is not 
conserved.\\
The stationary solution $n_s(\vec x)$ can be found from eq.(\ref{stat})
\begin{equation}
\mid \nabla n_s(\vec x) \mid = n_s (\vec x)^{2/(z-1)}
\end{equation}
For simplicity let us study a radial symmetric solution initiated by a 
radial symmetric initial condition. The exact result is  
\begin{equation}
n_s(r) = \left[c_1 + c \frac{z-3}{(2-d)(z-1)} r^{2-d} \right]^{(z-1)/(z-3)}
\end{equation} 
where $c$ and $c_1$ are constants. The special case $d=2$ is incorporated 
since $(2-d)^{-1} r^{2-d}$ leads to a logarithmic term
\begin{equation}
n_s(r) = (c_1 + c \ln r)^3
\end{equation}
For large distances the solution behaves like
\begin{equation}
n_s(r) \simeq c r^{-\alpha} \quad \mbox{with}\quad 
\alpha = \frac{(z-4)(z-1)}{2(z-3)} = \frac{(d-2)(2d-1)}{2d-3}
\label{rad}
\end{equation}
where for $d=2$ it results in $n_s \propto \ln^3 r$ which is different from conventional 
diffusion $n \propto \ln r$. Remark that the stationary solution in case of usual diffusion is 
described by an exponent $\alpha = d -2$. Therefore $d=2$ is a special case 
which will be also manifested in the dynamical analysis. \\
Now let us discuss the stability of the stationary solution against a time 
dependent perturbation.  To this aim we write $n(\vec x, t) = n_s(\vec x) 
+n_1(\vec x, t)$.  (From now on we include the coupling constant $J$ introduced in 
eq. (\ref{restr}) in the time variable $t$.) It is easy to check that the correction term $n_1$ 
obeys
\begin{equation}
\partial_t n_1 = (z-1)\left(n_1\nabla^2n_s + n_s \nabla^2n_1\right) - 
4\nabla n_s \nabla n_1
\label{sgl}
\end{equation}
Such a partial differential equation can be transformed into a diffusion like 
equation making the ansatz 
$n_1(\vec x,t) = \phi(\vec x) \psi(\vec x, t)$ where the function 
$\phi(\vec x)$ is chosen in such a manner that the linear gradient term 
disappears:
$$
(z-1)n_s \nabla \phi = 2 \phi \nabla n_s  
$$
As the result we find an equation where the contributions proportional to $\psi$ 
cancel each other. 
\begin{eqnarray}
\partial_t \psi(\vec x, t) &=& D(\vec x) \nabla^2 \psi(\vec x, t) \nonumber\\
\mbox{with}\quad D(\vec x) &=& (z-1) n_s(\vec x)
\label{sch}
\end{eqnarray}
As before let us study a radial symmetric solution. Taking into account 
eq.(\ref{rad}) and inspecting the last equation in comparison 
with the diffusion equation we make the ansatz
\begin{equation}
\psi \propto t^{-\gamma} \exp(-b\frac{r^{\alpha + 2}}{t})
\end{equation}
which allows to determine the unknown quantities $\gamma$ and $b$:
\begin{equation}
\gamma = \frac{\alpha + d}{\alpha + 2} \quad 
\mbox{and}\quad b = \frac{1}{c(z-1)(\alpha+2)^2}
\label{gam}
\end{equation}
Like before the result leads to the conventional behaviour for $d=2$. In the 
limit 
of an infinite range of the hopping process corresponding to $z \to \infty$ 
we find $\gamma = 2$.\\
Summarizing the results the radial symmetric solution of eq.(\ref{stat}) 
can be written for large distances $r$ in scaling form  
\begin{equation}
n(r,t) \simeq \frac{c}{r^{\alpha}} + t^{-\rho} f(\frac{r^{\alpha+2}}{t}) 
\label{dif}
\end{equation}
where the exponents and the scaling function are given by
\begin{equation}
\rho =   \frac{z-2}{\alpha+2 } \quad 
f(v)= A v^{(2 \alpha)/(\alpha+2)(z-1)} \exp(-b v)
\end{equation}
To derive the last equation we have also used the 
radial symmetric solution of the function $\phi(r)$ introduced after 
eq.(\ref{sgl}).\\   
Inspecting eq.(\ref{dif}) one remarks that the dynamical part of the 
solution yields a maximum. The position of this bump $r_b$ follows a simple 
law
\begin{equation}
r_b(t) \propto t^{\frac{1}{\alpha+2}}
\label{frac}
\end{equation}
Obviously, the bump is the more pronounced the lower the dimension. The result 
is depicted in Fig.1.\\
The exponent of diffusion $d_w$ is simply defined by eq.(\ref{frac}) 
\begin{equation}
d_w = \alpha + 2
\end{equation}
As mentioned the exponent agrees with that of conventional diffusion only 
in the two dimensional case. For instance for $d=3$ it results 
$d_w = \frac{11}{3}$. We anticipate that $d_w$ is considerably larger 
than $2$ since many of the sites are already occupied due to the restriction 
introduced by eq.(\ref{restr}). As the consequence a lot of the  
neighbouring sites of a given pair of sites are unavailable to be reached by 
particles. Instead of that they are obliged to return to the starting point. 
The restrictions act like a cage for the particles. A similar conclusion 
can be drawn in case of a fractal lattice which leads also to an 
exponent $d_w$ larger than two. The position of the bump $r_b$ 
moves much less rapidly with time $t$ than for an unrestricted hopping.\\
 
\noindent In the next step we would like to discuss the influence 
of a heat bath with a temperature $T$. To this aim the Fock space 
formulation is extended by replacing the evolution operator 
$\hat{L}$ by \cite{st4}
\begin{equation}
\hat{L} = \sum_{<ij>} J_{ij} 
\left[(1 - d_i d^{\dagger}_j) \exp(-H/2T) d^{\dagger}_i d_j \exp(H/2T) \right]
\label{di2}
\end{equation}
where the Hamiltonian will be chosen to be
\begin{equation}
H = h\sum_r n_r + K \sum_{<r.s>} n_r n_s
\end{equation}
Here $h$ means an external field (or chemical potential of the lattice gas) 
and $K$ stands for the static coupling between nearest neighbours situated at 
the cells $r$ and $s$. For $K > 0$ there is an additional repulsion whereas 
for $K < 0$ the particles are attractive. Remark that eq.\ (\ref{di2}) is in 
accordance with detailed balance.\\
Using the algebraic properties of Pauli--operators the evolution operator 
can be rewritten
\begin{eqnarray}
\hat{L} &=& \sum_{<ij>} J_{ij} 
\left[(1 - d_i d^{\dagger}_j) d^{\dagger}_i d_j A_i B_j \right] \nonumber\\
\mbox{with} \quad A_i &=& \exp\left(-K/T \sum_{s(i)} n_s\right) 
\quad \mbox{and} \quad B_i = A^{-1}_i
\end{eqnarray}
The exact evolution equation for the density reads now
\begin{eqnarray}
\partial_t \langle n_r \rangle = \exp(K/T) \sum_{j(r)}\langle J_{rj} 
(U_r \nabla^2 V_j - V_r \nabla^2 U_j ) \rangle \nonumber\\
\mbox{with} \quad U_r = (1 - n_r) A_r \quad \mbox{and} \quad V_r = n_r B_r
\end{eqnarray}
As far as the hopping parameter does not depend on the concentration itself 
the last equation can be represented as a continuity equation. However, 
in the case where the $J_{ij}$ are determined by the concentration cf. 
eq.(\ref{restr}) the conservation law is not valid as already remarked in the 
non--thermalized approach. After performing the continuum limit we obtain 
in the one dimensional case
\begin{equation}
\partial_t n = n \partial_{xx} n - 2 (\partial_x n)^2 + 
4 (K/T) (n \partial_{xx} n - (\partial_x n)^2  - n^2 \partial_{xx} n) 
\end{equation}
Applying the same procedure as before the last equation can be transformed in
\begin{eqnarray}
\partial_t \psi  &=& \tilde{D}(x) \partial_{xx} \psi\nonumber\\ 
\mbox{with}\qquad \tilde{D}(x) &=& n_s(x)(1+4 (K/T)n_s(x)(1-n_s(x))
\end{eqnarray}
The only difference to eq.(\ref{sch}) is that $D(x)$ is replaced by 
$\tilde{D}(x)$. If $K > 0$, i.e. repulsive interaction, the modified diffusion 
function $\tilde{D}(x)$ is always positive and the general behaviour is 
only slightly changed. 
However if $K < 0$ (attraction) the quantity $\tilde{D}(x)$ can be negative 
depending on the ratio $\frac{K}{T}$ and on the density $n_s$. A negative 
diffusion coefficient is obviously related to an instability.\\ 

\noindent In the present paper we have demonstrated that constraints during the 
diffusion process lead to a complete different behaviour. Due to the 
restrictions imposed on the diffusion process it appears a non--linear 
diffusion equation which can be analyzed partially in an analytical manner. 
There appears a maximum in the concentration which follows anomalous diffusion. 
Such a behaviour should be relevant for glassy materials characterized by 
rapid decrease of the mobility for decreasing temperatures. 
Further applications can be discussed when other degrees of freedom 
as impurities or vacancies are taken into account.

\begin{figure} 
\caption{3 dimensional plot $n(r,t)$ for $d=1$ } 
\label{fig1}
\end{figure}

\end{document}